# SOLAR NEUTRINO EXPERIMENTS: NEW PHYSICS?

JOHN N. BAHCALL

Institute for Advanced Study, Princeton, New Jersey 08540, USA

ABSTRACT

Physics beyond the simplest version of the standard electroweak model is required to reconcile the results of the chlorine and the Kamiokande solar neutrino experiments. None of the 1000 solar models in a full Monte Carlo simulation is consistent with the results of the chlorine or the Kamiokande experiments. Even if the solar models are forced artificially to have a $^8$B neutrino flux in agreement with the Kamiokande experiment, none of the fudged models agrees with the chlorine observations. This comparison shows that consistency of the chlorine and Kamiokande experiments requires some physical process that changes the shape of the $^8$B neutrino energy spectrum. The GALLEX and SAGE experiments, which currently have large statistical uncertainties, differ from the predictions of the standard solar model by $2\sigma$ and $3\sigma$, respectively. The possibility that the neutrino experiments are incorrect is briefly discussed.

## 1. New Physics is Required if Solar Neutrino Experiments are Correct.

Four solar neutrino experiments[1–4] yield results different from the combined predictions of the standard solar model and the standard electroweak model with zero neutrino masses. The question physicists most often ask each other about these results is: Do these experiments—if correct—require new physics beyond the standard electroweak model? For the purpose of this discussion, I shall mean by the standard electroweak model the simplest version in which neutrinos are massless and neutrino flavors are conserved.

Hans Bethe and I have argued[5] that new physics is required, that some process beyond the enormously successful electroweak model is required to change the shape of the $^8$B neutrino spectrum. I will review this argument below and then make some concluding remarks on whether the solution to the solar neutrino problem is more likely to be new physics or wrong experiments. The discussion will proceed by assuming that the standard electroweak model is correct, i.e., nothing happens to solar neutrinos while they are traveling to the earth from the point at which they are created in the sun. We shall see that this assumption leads us to a contradiction if the chlorine and Kamiokande solar neutrino experiments are correct.

The basis for our investigation is a collection of 1000 precise solar models[6] in which each input parameter (the principal nuclear reaction rates, the solar composition, the solar age, and the radiative opacity) for each model was drawn randomly from a normal distribution with the mean and standard deviation appropriate to that variable. The uncertainties in the neutrino cross sections for chlorine and for gallium were included by assuming a normal distribution for each of the absorption cross sections with its estimated mean and error.

This Monte Carlo study automatically takes account of the nonlinear relations among the different neutrino fluxes that are imposed by the coupled partial differential equations of stellar structure and by the boundary conditions of matching the observed solar luminosity, heavy element to hydrogen ratio, and effective temperature at the present solar age. This is the only Monte Carlo study to date that uses large numbers of standard solar models that satisfy the equations of stellar evolution. Just as detailed Monte Carlo calculations are necessary in order to understand the relative and absolute sensitivities of complicated laboratory experiments, a full Monte Carlo calculation is required to determine the interrelations and absolute values of the different solar neutrino fluxes. The sun is as complicated as a laboratory accelerator or a laboratory detector, for which we know by painful experience that detailed simulations are necessary. For example, the fact that the $^8$B flux may be crudely described as $\phi(^8B) \propto T^{18}_{\rm central}$ and $\phi(^7Be) \propto T^{8}_{\rm central}$ does not specify whether the two fluxes increase and decrease together or whether their changes are out of phase with each other. The actual variations of the calculated neutrino fluxes are determined by the coupled partial differential equations of stellar evolution and the boundary conditions, especially the constraint that the model luminosity at the present epoch be equal to the observed solar luminosity.

Figures 1a–c show the number of solar models with different predicted event rates for the chlorine solar neutrino experiment, the Kamiokande (neutrino-electron scattering) experiment, and the two gallium experiments (GALLEX and SAGE). For the chlorine experiment, which is sensitive to neutrinos above 0.8 Mev, the solar model with the best input parameters predicts[6] an event rate of about 8 SNU. None of the 1000 calculated solar models yields a capture rate below 5.8 SNU, while the observed rate is[1]

$$< \phi\sigma >_{\rm Cl\ exp} = (2.2 \pm 0.2)\ {\rm SNU}, \quad 1\sigma\ {\rm error}. \qquad (1)$$

The discrepancy that is apparent in Figure 1a was for two decades the entire "solar neutrino problem."

Figure 1a implies that something is wrong with either the standard solar model, with the chlorine experiment, or with the standard electroweak description of the neutrino. I assume in this part of the discussion that the solar neutrino experiments are correct with their indicated uncertainties. Therefore, we must try to decide here between the validity of the standard solar model and the standard electroweak theory. At first glance, the choice is clear: the standard electroweak model has been checked in exquisite detail by laboratory experiments; the standard solar model has tested extensively only in the outer regions in which no neutrinos are produced. "Surely," says our intuition, "the standard electroweak model must be correct and the standard solar model must be wrong." Not so, I think.

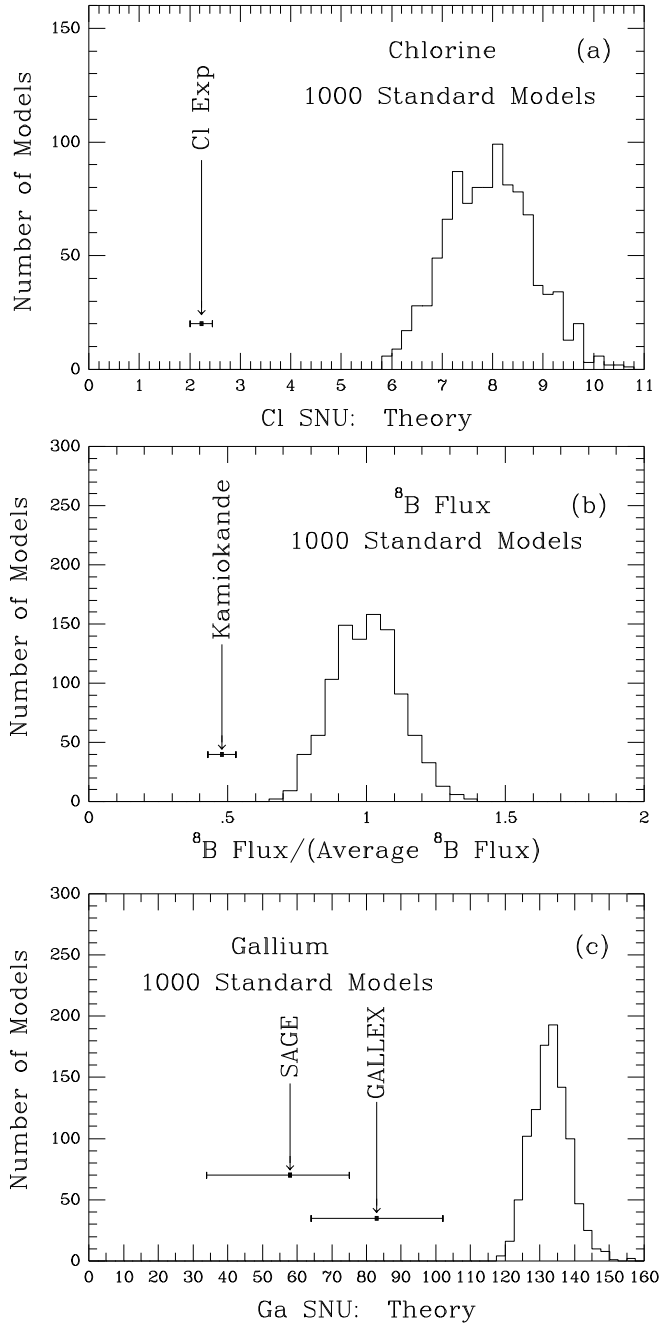

FIG. 1. 1000 Solar Models versus Experiments. The number of precisely-calculated solar models that predict different solar neutrino event rates are shown for the chlorine (Figure 1a), Kamiokande (Figure 1b), and gallium (Figure 1c) experiments. The solar models from which the fluxes were derived satisfy the equations of stellar evolution including the boundary conditions that the luminosity, chemical composition, and effective temperature at the current solar age be equal to the observed values. Each input parameter in each solar model was drawn independently from a normal distribution having the mean and the standard deviation appropriate to that parameter. The experimental error bars include both statistical errors ($1\sigma$) and systematic uncertainties, combined linearly.

Figure 1b shows the number of solar models with different $^8$B neutrino fluxes. For convenience, we have divided each $^8$B flux by the average $^8$B flux so that the distribution is peaked near 1.0. The rate measured for neutrinos with energies above 7.5 MeV by the Kamiokande II and III experiments is[2]

$$< \phi(^8B) > = [0.48 \pm 0.05(1\sigma) \pm 0.06(\text{syst})] < \phi(^8B) >_{\text{Average}} \qquad (2)$$

for recoil electrons with energies greater than 7.5 MeV. Here $< \phi(^8B) >_{\text{Average}}$ is the best-estimate theoretical prediction. None of the 1000 standard solar models lie below $0.65 < \phi(^8B) >_{\text{Average}}$. If one takes account of the Kamiokande measurement uncertainty($\pm 0.08$) in the Monte Carlo simulation, one still finds that none of the solar models are consistent with the observed event rate. This result provides independent support for the existence of a solar neutrino problem.

Figure 1c shows the number of solar models with different predicted event rates for gallium detectors and the recent measurements by the SAGE[3,7] ($58^{+17}_{-24} \pm 14(\text{syst})$ SNU) and the GALLEX ($83 \pm 19(1\sigma) \pm 8(\text{syst})$ SNU) collaborations[4]. With the current large statistical errors, the results differ from the best-estimate theoretical value[6] of 132 SNU by approximately 2 $\sigma$ (GALLEX) and 3.5 $\sigma$ (SAGE). The gallium results provide modest support for the existence of a solar neutrino problem, but by themselves do not constitute a strong conflict with standard theory.

Can the discrepancies between observation and calculation that are summarized in Figure 1 be resolved by changing some aspect of the solar model? This is difficult to do. A comparison of Figures 1a–1b shows that the discrepancy with theory is energy dependent. The larger discrepancy occurs for the chlorine experiment, which is sensitive to lower neutrino energies than is the Kamiokande experiment.

The key result to be used here is a detailed investigation—assuming no new physics beyond the standard electroweak model—which shows[8] that, to an accuracy of one part in $10^5$, the energy spectrum of the $^8$B solar neutrinos must have the same shape as the spectrum determined from laboratory nuclear physics experiments. The invariance of the energy spectrum allows us to compute the rate of neutrino capture in the chlorine experiment—independent of any considerations of solar models—provided only that we know from the Kamiokande experiment the flux of the higher energy (> 7.5 MeV) $^8$B neutrinos. In this process, we ignore the expected contributions to the chlorine experiment, which has a threshold of only 0.8 MeV (an order of magnitude less than the Kamiokande experiment), from $^7$Be, CNO, and pep neutrinos.

Using the empirical result obtained for the Kamiokande experiment[2], one finds that the predicted rate *from $^8$B neutrinos alone* in the chlorine experiment is 6.20 SNU (from the standard model)$\times 0.48$ (from the Kamiokande measurement), or

$$< \phi\sigma >_{\text{Cl; Kamiokande only}} =$$
$$[3.0 \pm 0.3(1\sigma) \pm 0.4(\text{syst})] \text{ SNU}. \qquad (3)$$

This minimum rate exceeds by more than $2\sigma$ the observed chlorine rate. The contributions of the other neutrino fluxes are more robustly calculated than the $^8$B flux which yields the minimum rate given above. Therefore, these other fluxes make more acute the disagreement between calculated and observed rates. However, one cannot apply to these other fluxes the same argument—based upon the invariance of the shape of the energy spectrum—given above since they are below threshold for the Kamiokande experiment.

Figure 2 is the central result upon which Hans and I based our conclusions. This figure provides a quantitative expression of the difficulty in reconciling the Kamiokande and chlorine experiments by changing solar physics. We constructed Figure 2 using the same 1000 solar models as were used in constructing Figure 1, but for Figure 2 we artificially replaced the $^8$B flux for each standard model by a value drawn randomly for that model from a normal distribution with the mean and the standard deviation measured by Kamiokande (see Eq. (2)). This *ad hoc* replacement is motivated by the fact that the $^7$Be $(p,\gamma)$ $^8$B cross section is the least accurately measured of all the relevant nuclear fusion cross sections and by the remark that the $^8$B neutrino flux is more sensitive to solar interior conditions than any of the other neutrino fluxes. The peak of the resulting distribution is moved to 4.7 SNU (from 8 SNU) and the full width of the peak is decreased by about a factor of three. The peak is displaced because the measured (i.e., Kamiokande) value of the $^8$B flux is smaller than the calculated value. The width of the distribution is decreased because the error in the Kamiokande measurement is less than the estimated theoretical uncertainty ($\approx 12.5\%$) and because $^8$B neutrinos constitute a smaller fraction of each displaced rate than of the corresponding standard rate.

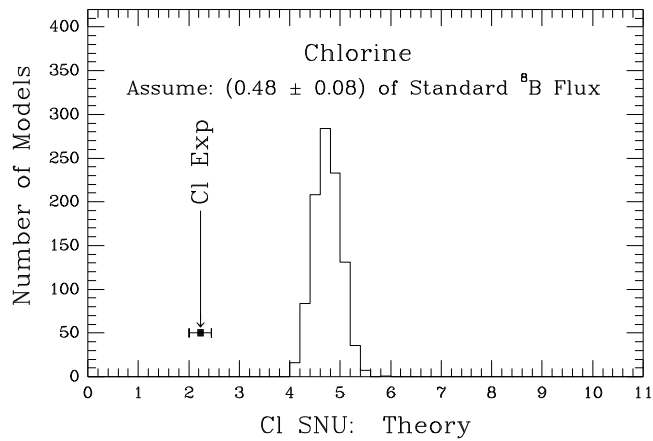

FIG. 2. 1000 Artificially Modified Fluxes. The $^8$B neutrino fluxes computed for the 1000 accurate solar models were replaced in the figure shown by values drawn randomly for each model from a normal distribution with the mean and the standard deviation measured by the Kamiokande experiment.

Figure 2 was constructed by assuming that something is seriously wrong with the standard solar model, something that is sufficient to cause the $^8$B flux to be reduced to the value measured in the Kamiokande experiment. Nevertheless, there is no overlap between the distribution of fudged standard model rates and the measured chlorine rate. None of the 1000 fudged models lie within $3\sigma$ (chlorine measurement errors) of the experimental result.

The results presented in Figures 1–2 suggest that new physics is required beyond the standard electroweak theory if the existing solar neutrino experiments are correct within their quoted uncertainties. Even if one abuses the solar models by artificially imposing consistency with the Kamiokande experiment, the resulting predictions of all 1000 of the "fudged" solar models are inconsistent with the result of the chlorine experiment(see Figure 2).

The difference between the minimum rate of 3.0 SNU for the chlorine experiment obtained by making use of the invariance of the $^8$B neutrino spectrum and the larger values shown in Figure 2 depends upon our understanding of the solar interior. In the future, it will be possible to use solar neutrinos to test electroweak theory independent of solar models by measuring the energy spectrum of the $^8$B neutrinos with the Super-Kamiokande[9], the SNO[10], and the ICARUS experiments[11] and by measuring the ratio of charged to neutral currents with the SNO experiment[10].

## 2. Are the Chlorine and Kamiokande Experiments Correct?

New solar neutrino experiments presently under development will decide if the current solar neutrino experiments are correct. The next generation of experiments, SNO, Super-Kamiokande, BOREXINO, and ICARUS, will settle this question. However, being a physicist at a festive occasion for physicists, I cannot resist summarizing my thoughts on the matter.

First of all, the strong reasons for caution. Neutrino experiments are fiendishly difficult. There have been a number of reported results of experiments with laboratory neutrinos that have—upon further investigation—subsequently required major revisions or outright rejection. Also, as a matter of historical record, we almost never learn new physics by trying to resolve astronomical puzzles. New physics has in the past come from the rigorously-defined conditions of laboratory experiments or from new theoretical formalisms, not from the mushier arguments of astrophysics.

Nevertheless, I think it is likely that the solution of the solar neutrino problem will be new physics rather than either revisions of the standard solar model or corrections to current solar neutrino experiments. Here are my reasons.

The predictions of the standard solar model are robust. The absolute values of the event rates for the chlorine experiment that I and my colleagues have calculated for the past 25 years are within the currently-quoted (modest) theoretical errors. This fact is illustrated in Figure 1.2 of the book *Neutrino Astrophysics*. Recent precise measurements of thousands of eigenfrequencies of the solar pressure waves are in essential agreement with predictions of the standard solar model[6]. Unfortunately, the fact that accurate measurements of the pressure modes are largely confined—at

present—to the outer half of the solar mass limits the significance, for our purposes, of confirmatory helioseismological measurements.

The chlorine and Kamiokande experiments disagree with each other independent of solar models. Assuming the invariance of the shape of the $^8$B neutrino energy spectrum that is promised by the standard electroweak model[8], the calculated rate for the $^8$B neutrinos that are observed in the Kamiokande experiment exceeds the observed rate in the chlorine experiment (see Eq. 3). This discrepancy is made much more severe when the contributions of the other, more robustly calculated neutrino fluxes are included in the predictions for the chlorine experiment (see Fig. 2).

The chlorine and Kamiokande experiments have been carefully examined by many non-affiliated researchers and have stood the test of time. They have also been subject to a variety of internal tests of self-consistency[1,2].

Finally, there is a simple and beautiful extension of the standard electroweak model that violates the conventional assumptions of zero neutrino masses and flavor conservation and does so in a way that solves the solar neutrino problem. The well-known MSW mechanism[12] can account for all the known solar neutrino experiments with a neutrino mass difference (of order 0.003 eV) that is consistent with current theoretical ideas regarding physics that might lie beyond the standard model[13]. Many other explanations of the solar neutrino problem that involve new, but apparently less "natural", physics have also been advanced.

I am willing to bet that the next generation of solar neutrino experiments will establish current suspicions that new physics is required to solve the solar neutrino problem. But, the wonderful thing about our subject is that in a few years we will not have to depend upon plausibility arguments; the experiments under development will determine the solution of the solar neutrino problem.

## 3. Acknowledgements

This work was supported in part by the NSF via grant PHY 92-45317 at I.A.S.